\documentclass[useAMS,usenatbib]{mn2e}

\usepackage{graphicx}
\usepackage{color}
\usepackage{amsmath}
\newcommand{\aap}{Astronomy and Astrophysics}
\newcommand{\jcap}{Journal of Cosmology and Astroparticle Physics}
\newcommand{\mnras}{Monthly Notices of the RAS}
\newcommand{\apj}{Astrophysical Journal}
\newcommand{\apjl}{Astrophysical Journal, Letters}
\newcommand{\apjs}{The Astrophysical Journal Supplement Series}
\newcommand{\aj}{The Astronomical Journal}
\newcommand{\pasp}{Publications of the Astronomical Society of the Pacific}

 \title[Predicted multiply-imaged X-ray AGNs in the XXL survey]{Predicted multiply-imaged X-ray AGNs in the XXL survey}

\author[F. Finet$^{1,2}$ et al.]{
F. Finet$^{1,2}$\thanks{E-mail:finet@astro.ulg.ac.be}, 
A. Elyiv$^{2,3,4}$,
O. Melnyk$^{2,5}$,
O. Wertz$^{2}$,
C. Horellou$^{6}$,
J. Surdej$^{2}$\thanks{Also, Directeur de Recherche honoraire du F.R.S.-FNRS}\\
$^{1}$Aryabhatta Research Institute of Observational Sciences (ARIES),
 			Manora Peak, Nainital-263 129, Uttarakhand (India)\\
$^{2}$Extragalactic Astrophysics and Space Observations (AEOS), University of Li\`ege,\\
         	 All\'ee du 6 Ao\^ut, 17 (Sart Tilman, B\^at. B5c), 4000 Li\`ege, Belgium\\
$^{3}$Main Astronomical Observatory, Academy of Sciences of Ukraine,
			27 Akademika Zabolotnoho St., 03680 Kyiv, Ukraine\\
$^{4}$Dipartimento di Fisica e Astronomia, Universit\`a di Bologna, 
			Viale Berti Pichat 6/2, I-40127  Bologna, Italy\\
$^{5}$Astronomical Observatory, Kyiv National University, 
		 	3 Observatorna St., 04053 Kyiv, Ukraine\\
$^{6}$Dept.of Earth \& Space Sciences, Chalmers University of Technology, Onsala Space Observatory, SE-439 92 Onsala, Sweden}

\begin{document}

\graphicspath{{images/}}

\date{Accepted yyyy Month dd. Received yyyy Month dd; in original form yyyy Month dd}

\pagerange{\pageref{firstpage}--\pageref{lastpage}} \pubyear{yyyy}

\maketitle

\label{firstpage}

\begin{abstract}
We estimate the incidence of multiply-imaged AGNs among the optical counterparts of X-ray selected point-like sources in the XXL field.
We also derive the expected statistical properties of this sample, such as the redshift distribution of the lensed sources and of the deflectors that lead to the formation of multiple images, modelling the deflectors using both spherical (SIS) and ellipsoidal (SIE) singular isothermal mass distributions.
We further assume that the XXL survey sample has the same overall properties as the smaller XMM-COSMOS sample restricted to the same flux limits and taking into account the detection probability of the XXL survey. 

Among the X-ray sources with a flux in the $\left [0.5-2\right ]$ keV band larger than $3.0 \times 10^{-15}$ erg cm$^{-2}$ s$^{-1}$ and with optical counterparts brighter than an $r$-band magnitude of 25, we expect $\sim 20$ multiply-imaged sources. Out of these, $\sim$16 should be detected if the search is made among the seeing-limited images of the X-ray AGN optical counterparts and only one of them should be composed of more than two lensed images. Finally, we study the impact of the cosmological model on the expected fraction of lensed sources.
\end{abstract}


\section{Introduction}

The XXL survey\footnote{http://ifru.cea.fr/xxl}, carried out by the space-based X-ray observatory XMM-Newton, spans over $\sim 2 \times 25$ square degrees with near 10 ks exposure in each field and 
is expected to lead to the detection of $\sim 25000$ Active Galactic Nuclei (AGNs) down to a limiting flux $10^{-15}$ erg cm$^{-2}$ s$^{-1}$ in the $\left [0.5-2\right ]$ keV soft X-ray band (\citealt{Pierre2015}). 
These X-ray data are complemented by multi-wavelength data obtained with the \textit{Canada-France-Hawaii Telescope Legacy Survey} (CFHTLS) and with the Blanco telescope (Blanco/South Pole Telescope (SPT) Cosmology Survey, BCS) in the (near-)optical $u$', $g$, $r$, $i$ and $z$ bands, down to a limiting AB magnitude of $\sim 25$.
Beside the multi-band imaging of the XXL fields, there is a very large on-going effort to obtain optical spectra of XXL sources, through either the matching of existing survey catalogues or dedicated spectroscopic surveys. Among these spectroscopic data acquisition programmes, the \textit{VIMOS Public Extragalactic Redshift Survey} (VIPERS, \citealt{Guzzo2010}) covers most of the northern field, the southern field being covered using the AAOmega multi-object spectrograph on the Anglo-Australian Telescope, an instrument used for the \textit{Galaxy and mass assembly} project (GAMA, \citealt{Driver2009}).

The completeness of this multi-wavelength database over the entire XXL field provides a unique sample to search for multiply-imaged AGNs. We have thus initiated such a search among the optical counterparts of point-like sources in the soft X-ray band. Beside the scientific interest provided by each multiply-imaged source, the goal of this project is to construct a statistically clean sample of lensed sources that will be used, in combination with samples of multiply-imaged sources from other recent surveys, to independently constrain the cosmological model.

The choice of the soft X-ray point-like sources is motivated by the higher sensitivity of XMM-Newton in this band.
Furthermore, this spectral band should contain a larger fraction of type-I AGNs than the hard X-ray. On average, type-I AGNs with a detectable optical counterpart are expected to have a higher redshift than type-II AGNs (more absorbed in the visible and thus more difficult to detect in the optical at high redshift). As higher redshift sources have a higher probability of being lensed, this population is more likely to undergo gravitational lensing with the formation of multiple images.  
The better angular resolution achievable in the optical domain will allow to unravel the multiply-imaged sources.

A search for gravitational lenses among the optical counterparts of X-ray sources has already been carried out for a subset of the XXL field, the \textit{XMM-Large Scale Structure field} (XMM-LSS, \citealt{Elyiv2013}). For this smaller field, visual inspection of all optical counterparts has been done in order to identify the multiply-imaged AGN candidates that are now awaiting spectroscopic confirmation. The search for lensed sources in the larger XXL field is in progress.

In this paper, we present a prospective analysis of the lensed AGN population detectable within the XXL survey, as well as a study of their expected statistical properties. In order to perform this analysis, we have reformulated the  mathematical formalism to study the statistical aspects of gravitational lensing, basing the statistical formalism on the observables of the source population.

This paper is structured as follows. We present the mathematical approach in Section \ref{section_theory}. Namely, we derive the expression allowing to calculate the probability for a source to be lensed with the formation of multiple images, modelling the deflector population by means of a spherical mass distribution and then taking into account the internal ellipticity of the deflector mass distribution. We explain how this expression may be averaged over the entire population of sources detected in the survey, thanks to the source joint probability density, with which we derive the expression of the expected fraction of lensed sources in a survey, as well as the expected redshift distribution of the lensed sources and that of the deflectors. Our simulations also account for the inability of the ground-based CFHT and Blanco telescopes to resolve multiple images with too small angular separations. 

In Section \ref{section_obs_constraints}, we present the observational constraints used to estimate the expected properties of the XXL population in the X-ray and optical domains: these were inferred from the deeper (but smaller) XMM-COSMOS field (\citealt{Brusa2010}). Finally, in Section \ref{section_results} we present our results, i.e. the expected number of multiply-imaged sources in the XXL and the XMM-LSS fields, as well as the expected statistical properties of these lensed sources. We also investigate the fraction of lensing events as a function of their number of lensed images and we investigate how the fraction of multiply-imaged sources changes as a function of the cosmological mass density parameter, $\Omega_{m}$.


\section{Mathematical formalism}
\label{section_theory}


\subsection{Lensing optical depth}

Multiple images due to gravitational lensing occur when light rays emitted from a background source are deviated towards the observer by a foreground deflector located near the line of sight. 
In our case, the amplified lensed sources have to be above the survey flux limit in both the X-ray and the optical domains, and the lensed images have to be resolved in the latter. The probability for a source to be lensed depends on its redshift, its X-ray flux and $r$-band magnitude. In this section, we derive an expression to calculate the probability for a source to undergo a gravitational lensing event as a function of its redshift and its X-ray flux only, considering the minimal angular separation resovable in the $r$-band. We will analyse the validity of this simplification in Section \ref{section_rbandCutoff} by formally including the $r$-magnitude in the calculations.

Let us consider a source with redshift $z_{s}$, with an observed flux $f_{X}$ in the X-ray band, and a lens with a mass distribution characterised by a set of parameters $M'$, located at an intermediate position along the line of sight at redshift $z_{d}$. In the lens plane perpendicular to the line of sight,  we can define an area $\Sigma\left (z_{s},z_{d},f_{X},M'\right )$ centred on the source projected on the lens plane, called the lensing cross section in which the presence of a deflector leads to the detection of multiple images by the observer (where the multiple images are resolved in the $r$-band CCD frames).
The lensing cross section is a function of the redshifts of the source and the deflector, the X-ray flux $f_{X}$ of the source and the deflector mass distribution parameters $M'$ (some mass distributions are more efficient at deflecting light rays and thus have a larger lensing cross section).

The probability $d\tau\left (z_{s}, z_{d}, f_{X},M'\right )$ for this source to be multiply imaged due to the presence of a deflector in the redshift range $\left [z_{d}, z_{d} + dz_{d}\right ]$ is given by (\citealt{Turner1984})
\begin{equation}
d\tau =  \left (1+z_{d}\right )^{3} n_{D}\left (z_{d},M'\right ) \, \frac{cdt}{dz_{d}} \,\Sigma\left (z_{s}, z_{d}, f_{X},M'\right ) \,  dz_{d},
\label{eq_def_dtGL}
\end{equation}
where $n_{D}\left (z_{d},M'\right )$  is the volume density in the comoving reference frame of deflectors characterised by the mass parameters $M'$. The quantity $cdt/dz_{d}$ represents the infinitesimal light-distance element at redshift $z_{d}$ per deflector redshift unit, which, in a flat expanding FLRW universe, is given by (e.g. \citealt{Peebles1993})
\begin{equation}
\frac{cdt}{dz_{d}} = \frac{c}{H_{0}\left (1+z_{d}\right )}\, \left [\left (1+z_{d}\right )^{3} \Omega_{m} + \left (1-\Omega_{m}\right )\right ]^{-1/2},
\label{eq_def_cdtdz}
\end{equation}
where $\Omega_{m}$ is the present-day value of the cosmological mass density parameter.

The envelope of the lensing cross sections at different deflector redshifts $z_{d}$ defines the \textit{lensing volume} in which the presence of a deflector leads to the detection of multiple lensed images of the background source.
The probability $\tau\left (z_{s},f_{X},M'\right )$ for a source to be lensed with the formation of multiple images can be calculated by integrating Eq. \ref{eq_def_dtGL} over all values of the deflector redshift $z_{d}$, which leads to
\begin{equation}
\begin{split}
&\tau\left (z_{s}, f_{X},M'\right ) =\\ 
&\int\limits_{0}^{z_{s}}\! \left (1+z_{d}\right )^{3} n_{D}\left (z_{d},M'\right )  \frac{cdt}{dz_{d}}\, \Sigma\left (z_{s},z_{d}, f_{X},M'\right ) \,  dz_{d}.
\end{split}
\label{eq_def_tGL}
\end{equation}
The definition of $\tau\left (z_{s}, f_{X},M'\right )$ in Eq. \ref{eq_def_tGL} corresponds to an optical depth, which for small values  can be assimilated to a probability. For this reason we refer to $\tau\left (z_{s}, f_{X},M'\right )$ as the source \textit{lensing optical depth} or \textit{lensing probability}, without distinction.

In Eq. \ref{eq_def_tGL}, the comoving density $n_{D}\left (z_{d},M'\right )$ of deflectors assumes one type of deflectors with similar characteristics defined by the parameters $M'$. Considering the mass distribution to be characterised by means of deflectors with a central velocity dispersion $\sigma$, in the range $\left [\sigma, \sigma + d\sigma\right ]$, $n_{D}\left (z_{d},\sigma\right )$ may be expressed by means of the \textit{Velocity Dispersion Function} (VDF) of galaxies $\Phi_{\sigma}\left (\sigma,z_{d}\right )$
\begin{equation}
n_{D}\left (z_{d},\sigma\right ) = \Phi_{\sigma}\left (\sigma,z_{d}\right) d\sigma.
\label{eq_def_nEG}
\end{equation}
To take into account the contribution of all galaxies with different central velocity dispersions, Eq. \ref{eq_def_tGL} has to be integrated over $\sigma$.
The VDF in Eq. \ref{eq_def_nEG} can be either measured directly or inferred from the Luminosity Function of the deflecting galaxies, using the mean Faber-Jackson or Tully-Fischer relationship, depending on the type of galaxies considered. However, \citet{Sheth2003} have shown that neglecting the dispersion of the Faber-Jackson relationship leads to a wrong estimate of the VDF. We thus use the VDF determined directly through observations (thereby following \citealt{Sheth2003}, \citealt{Mitchell2005}, \citealt{Choi2007} and \citealt{Chae2010}).
 \cite{Sheth2003} and \cite{Choi2007} have shown that the VDF of early- and late-type galaxies is well fitted by the modified Schechter function 
\begin{equation}
\Phi_{\sigma}\left (\sigma,z_{d}\right) \, d\sigma= \Phi_{\ast} \left (\frac{\sigma}{\sigma_{\ast}}\right )^{\alpha} \exp\left ( - \left (\frac{\sigma}{\sigma_{\ast}}\right )^{\beta}\right ) \frac{\beta}{\Gamma\left (\alpha/\beta\right )} \frac{d\sigma}{\sigma},
\label{eq_def_VDF}
\end{equation}
where $\Phi_{\ast}$ and $\sigma_{\ast}$ are the characteristic number density and central velocity dispersion, respectively, $\alpha$ and $\beta$ are the slope coefficients of the VDF for low and high values of $\sigma$, respectively, and where $\Gamma \left (x\right )$ is the Gamma function.

Although less numerous, early-type galaxies are much more efficient deflectors than late-type ones, which tend to form multiple images with smaller angular separation.
Late-type galaxies contribute by less than 10\% to the gravitational lensing events in a typical sample of lensed AGNs selected in the optical or near infrared (\citealt{Fukugita1991}, \citet{Maoz1993}, \citet{Keeton1998}, \cite{Kochanek2000}), although in radio-selected samples, thanks to the better angular resolution of the survey, the fraction of lensing events formed by late-type deflectors may be higher (e.g. the CLASS survey where at least 5 of 22 deflectors are late-type galaxies, see \citealt{Browne2003}). In the present work, because the multiple images will be searched for in the SDSS $r$-band, we consider the deflector population to be only composed of early-type galaxies, and we will study in Section \ref{section_results} the validity of this assumption.

Constraints from strong-lensing statistics on the evolution of the VDF of early-type galaxies show very little evolution or are consistent with a no-evolution assumption (see e.g. \citealt{Chae2003}, \citealt{Ofek2003}, \citealt{Chae2010} and \citealt{Oguri2012}). 
Consequently, throughout this work we assume the deflector VDF to be constant with redshift in the comoving reference frame, or in other words, that there is no impact of the evolution of the deflector population on the VDF and we use the value of the VDF parameters determined in the local universe by \citet{Choi2007}, i.e. $$\left [\Phi_{\ast},\sigma_{\ast},\alpha, \beta \right ] = \left [8\times 10^{-3} \,\mathrm{h}^{3}\, \mathrm{Mpc}^{-3},161 \,\mathrm{km}\, \mathrm{s}^{-1}, 2.32, 2.67 \right ].$$

The lensing cross section in Eq. \ref{eq_def_dtGL} depends on the deflector mass distribution parmeters $M'$.
For a deflector mass distribution modelled as a \textit{Singular Isothermal Sphere} (SIS) (a spherical mass distribution with a volume density scaling as $r^{-2}$), the mass distribution is characterised by the line of sight velocity dispersion $\sigma$ and the lensing cross section $\Sigma\left (z_{s},z_{d}, f_{X},\sigma\right )$ can be defined as (see \citealt{Turner1984} for definition and \citealt{Claeskens1999} for the formalism followed in this paper)
\begin{equation}
\Sigma\left (z_{s},z_{d},f_{X},\sigma\right ) = b_{0}^{2}\left (z_{s},z_{d}, \sigma\right ) \iint\limits_{S_{y}} \frac{N_{f_{X}}\left (f_{X}/A\left (\mathbf{y}\right )\right )}{N_{f_{X}}\left (f_{X}\right )} d\mathbf{y},
\label{eq_def_LCS}
\end{equation}
where we have introduced the source vector $\mathbf{y} = \left (y_{1},y_{2}\right )$ which Cartesian coordinates are projected on the deflector plane, and normalised to the scale factor $b_{0}$ (i.e. the Einstein radius in the deflector plane). 
$A\left (\mathbf{y}\right )$ is the total amplification of the lensed images formed for a source located at the position $\mathbf{y}$, i.e. the sum of the amplification moduli of each lensed image and $N_{f_{X}}\left (f_{X}\right )$ is the \textit{differential number counts function} (DNCF) as a function of the source flux $f_{X}$ in the X-ray band.
For an SIS deflector, the scale factor is given by (see e.g. \citealt{Claeskens1999})
\begin{equation}
b_{0}\left (\sigma,z_{d},z_{s},UM\right ) = 4 \pi \left (\frac{\sigma}{c}\right )^{2} \frac{D_{OD} D_{DS}}{D_{OS}},
\label{eq_def_b0}
\end{equation} 
where $D_{OD}$, $D_{DS}$ and $D_{OS}$ represent the different angular diameter distances between the observer, the deflector and the source, $c$ is the speed of light, and $UM = \left (\Omega_{m}, H_{0}\right )$ is a set of parameter values characterising the universe model (UM) as a flat expanding FLRW one.

The ratio $N_{f_{X}}\left (f_{X}/A\left (\mathbf{y}\right )\right )/N_{f_{X}}\left (f_{X}\right )$ in Eq. \ref{eq_def_LCS} is known as the \textit{amplification bias} (\citealt{Turner1980}, \citealt{Turner1984}, \citealt{Fukugita1991}). It is introduced to take into account a favourable bias when estimating the lensing probability in a flux-limited sample induced by the amplification phenomenon. The amplification may lead to the inclusion of sources in a flux-limited sample that are intrinsically fainter than the flux limit but have undergone a gravitational lensing amplification. 
Since the ratio of the DNCF in Eq. \ref{eq_def_LCS} at two different X-ray flux levels (i.e. $f_{X}$ and $f_{X}/A\left (\mathbf{y}\right )$) is likely to be slightly dependent on the source redshift, it would certainly be more accurate to consider the redshift dependence of the DNCF. 
However, at the time of this work, the XXL redshift are not yet available and we do not have access to a source sample large enough to constrain the redshift dependence of the DNCF (the source sample used is described in Section \ref{section_obs_constraints}).
Consequently, we consider the value of the DNCF ratios averaged over the whole redshift range.
Because of the presence of $A\left (\mathbf{y}\right )$ in the amplification bias, the expression of $\Sigma\left (z_{s},z_{d}, f_{X},\sigma\right )$ is intrinsically linked to the amplification map of the deflector defining the amplification as a function of the source position, itself determined by the mass distribution of the deflector. It is out of the scope of this paper to explain this dependence in detail and the reader may refer to e.g. \cite{Hezaveh2011} for some further description. Nevertheless, because of the presence of the total amplification $A$ in its argument, the calculation of the amplification bias in Eq. \ref{eq_def_LCS} necessitates the knowledge of $N_{f_{X}}\left (f_{X}\right )$  for sources fainter than the survey limiting flux. This may be estimated either by extrapolating $N_{f_{X}}\left (f_{X}\right )$ or by considering a parent source sample accessing fainter fluxes. In this work, we will consider a deeper source sample described in Section \ref{section_obs_constraints}.

Finally, $S_{y}$ in Eq. \ref{eq_def_LCS} represents the area in the $\mathbf{y}$ plane, centred on the deflector, in which the presence of a source leads to a \textit{lensing event}, i.e. the lensing cross section, normalised to the scale factor. 

The \textit{lensing event} may be defined in different ways. It may refer to the formation of multiple images or to the formation of a given number of lensed images, or the formation of lensed images with an angular separation sufficiently large to be detected in a survey. Depending on the definition adopted for the lensing event, the integration domain $S_{y}$ differs and, consequently, this leads to a different definition of the lensing cross section $\Sigma$ in Eq. \ref{eq_def_LCS} and of the lensing optical depth in Eq. \ref{eq_def_tGL}. In most cases, the integration in Eq. \ref{eq_def_LCS} must be performed numerically.

When modelling the deflector mass distribution by means of an SIS profile, the lensing events can lead to the formation of two images at maximum. We define $\tau_{SIS}\left (z_{s},f_{X}\right )$ as the probability for a source to be multiply imaged when the deflectors are modelled by an SIS mass distribution. $\tau_{SIS}\left (z_{s},f_{X}\right )$ is thus calculated by inserting Eqs. \ref{eq_def_nEG}, \ref{eq_def_VDF}, \ref{eq_def_LCS} and \ref{eq_def_b0} into Eq. \ref{eq_def_tGL} and by performing the integration over $\sigma$. The integration over $\sigma$ can be performed analytically, leading to the following expression (\citealt{Turner1984}, or \citealt{Mitchell2005} for derivation using the VDF)
\begin{equation}
\begin{split}
&\tau_{SIS}\left (z_{s}, f_{X}\right ) = 
\Phi_{\ast} \frac{\Gamma\left (\left (\alpha + 4\right )/\beta\right )}{\Gamma\left (\alpha / \beta\right )}\\
&\int_{0}^{z_{s}} \left (1+z_{d}\right )^{3}\frac{cdt}{dz_{d}} \Sigma_{SIS}\left (z_{s}, z_{d}, f_{X},\sigma_{\ast}\right ) \,  dz_{d}.\\
\end{split}
\label{eq_t_SIS}
\end{equation}
$\Sigma_{SIS}\left (z_{s}, z_{d}, f_{X},\sigma_{\ast}\right )$ represents the lensing cross section defined by Eq. \ref{eq_def_LCS}, for $\sigma = \sigma_{\ast}$, when considering the area $S_{y}$ for the case of an SIS deflector. The integration in Eq. \ref{eq_t_SIS} must be performed numerically.

The introduction of an internal ellipticity in the mass distribution used to model the deflectors, allows to account for the formation of more than two lensed images. This is the reason why the \textit{Singular Isothermal Ellipsoid} (SIE) mass distribution has been introduced (\citealt{Kormann1994}). The ellipticity parameter in the SIE profile is the axis ratio $q$ of the deflector mass projected on the deflector plane and the mass distribution parameters $M'$ are now $\sigma$ and $q$. As for the case of the SIS deflector, we can define the probability $\tau_{SIE}\left (z_{s},f_{X}\right )$ for a source to be lensed with the formation of multiple images, irrespectively of the number of the lensed images, when modelling the deflector population with SIE mass distributions (\citealt{Huterer2005}). On the other hand, we may also define the probability $\tau_{SIE,i}\left (z_{s},f_{X}\right )$ for a source to be lensed with the formation of $i$ images (with $i=2,3\, \mathrm{ or }\, 4$).
As the lensing cross section now depends on the axis ratio $q$ of the deflector (through the dependence of $S_{y}$ in Eq. \ref{eq_def_LCS}), in order to calculate $\tau_{SIE}\left (z_{s},f_{X}\right )$ we must also integrate Eq. \ref{eq_def_dtGL} over the axis ratio $q$, using an appropriate probability distribution. Furthermore, the deflector density function in Eqs. \ref{eq_def_tGL} and \ref{eq_def_nEG} must take into account how the deflector population is distributed as a function of both $\sigma$ and $q$, and thus we have to introduce the dependence of $n_{D}$ on the axis ratio $q$. In Eq. \ref{eq_def_nEG}, the density $n_{D}\left (\sigma,q\right )$ of deflectors with a central velocity dispersion in the range $\left [\sigma, \sigma + d\sigma \right [$ and an axis ratio in the range $\left [q, q + dq \right [$ may be expressed as
\begin{equation}
\begin{split}
n_{D}\left (\sigma,q\right) &= \Phi_{\sigma}\left (\sigma\right) \, d_{q|\sigma}\left (\sigma,q\right) \,d\sigma \\
&= \Phi_{\sigma}\left (\sigma\right) \, d_{q}\left (q\right) \,d\sigma dq,
\label{eq_def_f_dist}
\end{split}
\end{equation}
where $d_{q|\sigma}\left (\sigma,q\right)$ represents the normalised axis ratio distribution for the deflectors with a central velocity dispersion $\sigma$, and $d_{q}\left (q\right)$ is the marginal normalised distribution as a function of $q$. The last equality arises if we assume that the deflector distributions as a function of $\sigma$ and $q$ are mutually independent\footnote{Although not strictly justified, this assumption is made because of the lack of observational constraints for the distribution of early-type galaxies in the $\left (\sigma,q\right )$ plane, as well as for calculation time consideration.}.
 $\Phi_{\sigma}\left (\sigma\right)$ is given by Eq. \ref{eq_def_VDF}, as in the case of the SIS mass model. There are strong evidences from the study of various gravitational lens samples, that elliptical galaxy isophotes and the mass distribution ellipticities are aligned and have well correlated values (see \citealt{Koopmans2006} and \citealt{Sluse2012} for independent confirmations). Furthermore, as previously mentioned, there is no evidence for strong evolution effects in the VDF of early-type galaxies from lensing surveys. Thanks to these two observational facts, $d_{q}\left (q\right)$ can be estimated from the distribution of the isophotes of early-type galaxies as measured in the local universe. We therefore use the axis ratio distribution measured by \cite{Choi2007} from a sample of elliptical galaxies in the local universe.

$\tau_{SIE}\left (z_{s}, f_{X}\right )$ can thus be calculated by inserting Eqs. \ref{eq_def_VDF} and \ref{eq_def_f_dist} into Eq. \ref{eq_def_dtGL}, and by integrating over $\sigma$ and $q$. Using Eqs. \ref{eq_def_LCS} and \ref{eq_def_b0}, and performing the analytical integration over $\sigma$, $\tau_{SIE}$ can thus be expressed as
\begin{equation}
\begin{split}
&\tau_{SIE}\left (z_{s}, f_{X}\right ) = \Phi_{\ast} \frac{\Gamma\left (\left (\alpha + 4\right )/\beta\right )}{\Gamma\left (\alpha / \beta\right )} \\
&\int\limits_{0}^{z_{s}}\!\left (1+z_{d}\right )^{3}\frac{cdt}{dz_{d}} \int\limits_{0}^{1}\!  d_{q}\left (q\right )\Sigma_{SIE}\left (z_{s},z_{d},f_{X},\sigma_{\ast},q\right ) \,  dq \,dz_{d},\\
\end{split}
\label{eq_t_SIE}
\end{equation}
where $\Sigma_{SIE}\left (z_{s},z_{d},f_{X},\sigma_{\ast},q\right )$ represents the lensing cross section calculated through numerical integration of Eq. \ref{eq_def_LCS}, when considering an SIE deflector with an axis ratio $q$ and a central velocity dispersion $\sigma_{\ast}$. Similarly, the probability $\tau_{SIE,i}\left (z_{s},f_{X}\right )$ for a source to be lensed with the formation of $i$ images, when modelling the deflectors with SIE profiles, can be calculated using $\Sigma_{SIE,i}$ in the previous relation and considering in $S_{y}$ only the area in which a source should be located in order to lead to the formation of $i$ lensed images.

We have developed \textit{Matlab} toolboxes and libraries to perform the numerical integration in the expressions of $\tau_{SIS}\left (z_{s},f_{X}\right )$ and $\tau_{SIE}\left (z_{s},f_{X}\right )$ from Eqs. \ref{eq_t_SIS} and \ref{eq_t_SIE}, as well as for the calculation of $\tau_{SIE,i}\left (z_{s},f_{X}\right )$, taking into account that lensed images angularly too close to each other cannot be resolved in the survey. 
The numerical integrations are made in two steps. First, we create a data base of the lensing cross sections in Eq. \ref{eq_def_LCS}, considering $b_{0}=1$, ranging over all possible values of $f_{X}$ and over the ratio $\theta_{mis}/\theta_{E}$, where $\theta_{mis}$ represents the smallest angular separation for which point-like images of same brightness are resolved in the survey, and also over $q$ for the SIE case. The integration of the double integral in Eq. \ref{eq_def_LCS} is performed by plain Monte-Carlo integration, where we randomly generate $\sim 10^{6}$ source positions $\mathbf{y}$, calculate the position and amplification of the lensed images, which contribute to the integral only if the multiple images can be resolved and brighter than the X-ray limiting flux of the survey.
In the second step, the lensing optical depths are calculated by integrating Eqs. \ref{eq_t_SIS} and \ref{eq_t_SIE} using trapezoidal integration and the database of lensing cross sections.
We have thus all the tools needed for the calculation of the lensing optical depth of a source with known redshift and apparent X-ray flux, when considering a population of deflectors modelled with SIS or SIE mass distributions.

Similar expressions for the lensing optical depths such as $\tau_{SIS}\left (z_{s},f_{X}\right )$, $\tau_{SIE}\left (z_{s},f_{X}\right )$ and $\tau_{SIE,i}\left (z_{s},f_{X}\right )$ (Eqs. \ref{eq_t_SIS} and \ref{eq_t_SIE}, respectively), as well as their differential contribution as a function of the deflector redshift $d\tau/dz_{d}$, have already been derived and used for the analysis of statistical samples of lensed sources, either to constrain the cosmological model (e.g. \citealt{Turner1984}, \citealt{Turner1990}, \citealt{Fukugita1990}, \citealt{Fukugita1991}, \citealt{Kochanek1992}, \citealt{Surdej1993}, \citealt{Maoz1993}, \citealt{Cen1994}, \citealt{Keeton2002}, \citealt{Chae2002}, \citealt{Chae2003}, \citealt{Oguri2012}),
 or to study the population of deflectors (\citealt{Keeton1998}, \citealt{Keeton1998a}, \citealt{Keeton1998b}, \citealt{Kochanek2000}, \citealt{Ofek2003}, \citealt{Chae2003}, \citealt{Chae2010}), some of the work having considered the ellipticity of the mass distribution as well as the effect of external sheer on the statistics (\citealt{Huterer2005}, \citealt{Oguri2010}, \citealt{Oguri2012}).

The various expressions derived for the optical depth $\tau$ and its differential contribution $d\tau/dz_{d}$ are considered for a single source. This may be averaged over the whole detected population to derive the mean optical depth through the sample as well as the expected redshift distributions of the lensed sources and deflectors. To average over the population of sources, some previous works have made use of the source luminosity function, and integrate over the absolute magnitude of the source population (e.g. \citealt{Oguri2010}, \citealt{Oguri2012}) which, for the integration process, necessitates the choice of a universe model. 

In the next subsection, we propose a slightly different formulation that allows to average any function over the entire population of sources detected in a survey, where the averaging is done based on the observables, i.e. the distribution in the $\left (z_{s},f_{X}\right )$ plane of the detected sources. Because the formalism is based directly on the observed distribution of sources, it naturally takes into account the detection biases of the sources. 
We apply this method to derive useful expressions such as the average lensing optical depth in a sample, the redshift distributions of the lensed sources and of the deflectors effectively leading to the formation of multiple lensed images.


\subsection{Joint source probability density and fraction of multiply-imaged sources}


Let us consider a survey characterised by its limiting flux, different biases in the source detection procedure and its angular coverage. Each detected source is characterised by its redshift $z_{s}$ and its flux $f_{X}$ in the selected spectral band (in the present case, the $\left [0.5 - 2\right ]$ keV band). The detection of a source within the survey may be considered as a random event with respect to the continuous random variables associated with the source redshift and the X-ray flux, respectively. 
	
We can define a probability $P\left (z_{s}, f_{X}\right )$ that a source detected in the survey is characterised by an observed redshift and flux in the range $\left [ z_{s}, z_{s} + dz_{s}\right ]$ and $\left [ f_{X}, f_{X} + df_{X}\right ]$, respectively. 
We may then define the joint probability density $d_{obs}\left (z_{s},f_{X}\right )$ spanning over the $\left (z_{s},f_{X}\right )$ plane, associated with this random event. $P\left (z_{s},f_{X}\right )$ and $d_{obs}\left (z_{s},f_{X}\right )$ are related through
\begin{equation} 
P\left (z_{s},f_{X}\right ) = d_{obs}\left (z_{s},f_{X}\right )  dz_{s} df_{X}  .
\end{equation}
The random variables $z_{s}$ and $f_{X}$ associated with a detected source follow the joint distribution described by $d_{obs}\left (z_{s},f_{X}\right )$. This function contains all the information about the survey and implicitly takes into account the detection biases. For a sufficiently large number of detected sources, the joint probability density $d_{obs}\left (z_{s},f_{X}\right )$ may be directly estimated from the detected source population, by calculating a smoothed histogram of the source distribution in the $\left (z_{s},f_{X}\right )$ plane, normalised by the total number $N_{AGN}$ of sources detected within the survey.
Ideally, we would have liked to define the source joint probability density $d_{obs}\left (z_{s},f_{X},r\right )$ in the $\left (z_{s},f_{X},r\right )$ space, where $r$ represents the SDSS $r$-magnitude of the optical counterpart(s) of the X-ray sources. However, because of the small number of sources observed in our reference sample (see Section \ref{section_obs_constraints}), the quantity $d_{obs}\left (z_{s},f_{X},r\right )$ could hardly be accurately determined. When analysing the XXL sample however, the number of detected sources should be sufficient to constrain $d_{obs}$ in the 3D-space. We shall therefore postpone such a more detailed study until all optical counterparts of the XXL X-ray sources have been identified.

The normalised marginal probability density distributions associated with the random variables $z_{s}$ and $f_{X}$ are closely related to the observations. Indeed, the marginal density distribution obtained by integating $d_{obs}\left (z_{s},f_{X}\right )$ over $z_{s}$ or $f_{X}$ represent the normalised source distribution as a function of the flux $N_{f_{X}}\left (f_{X}\right )/N_{AGN}$ and the redshift $N_{z_{s}}\left (z_{s}\right )/N_{AGN}$, respectively.

The use of $d_{obs}\left (z_{s},f_{X}\right )$ as a weighing function when performing the integration over the entire population of detected sources allows to calculate the expected mean value of any function of the random variables $z_{s}$ and $f_{X}$.
The mathematical expectation $\left < \tau \right >$ of the lensing optical depth, i.e. the fraction of sources gravitationally lensed within the detected population, can be calculated
by integating $\tau\left (z_{s},f_{X}\right )$ over the $\left (z_{s},f_{X}\right )$ plane, weighing with $d_{obs}\left (z_{s},f_{X}\right )$, i.e. 
\begin{equation}
\left < \tau \right > = \iint  \tau\left (z_{s},f_{X}\right ) d_{obs}\left (z_{s},f_{X}\right ) dz_{s} df_{X}. 
\label{eq_def_mean_tGL}
\end{equation}
Using the expression of $\tau_{SIS}\left (z_{s},f_{X}\right )$, $\tau_{SIE}\left (z_{s},f_{X}\right )$ or $\tau_{SIE,i}\left (z_{s},f_{X}\right )$ given by Eqs. \ref{eq_t_SIS} and \ref{eq_t_SIE} for the calculation of the optical depth $\tau\left (z_{s},f_{X}\right )$ in Eq. \ref{eq_def_mean_tGL}, we are able to calculate the expected fraction of multiply-imaged sources, considering a population of deflectors modelled with either the SIS or SIE mass profile. In the latter case, we can also calculate the expected fraction of lensed sources as a function of the number $i$ of lensed images.

\cite{Oguri2008} have also derived an expression for the number of multiply-imaged sources using a binning of the redshift-magnitude space (Eq. 12 in \citealt{Oguri2008}), and with the number of sources in the bins as a weighing factor. Their expression corresponds to the discrete equivalent of Eq. \ref{eq_def_mean_tGL}, integrating in the redshift-magnitude space, rather than the $\left (z_{s},f_{X}\right )$ plane. The essential difference with the approach used in these previous works and ours is the use of the source distribution in the $\left (f_{X},z_{s}\right )$ rather than using the source luminosity function as a weighing factor in the absolute magnitude-redshift space. Our approach allows to account directly for the detection bias of the sources and does not necessitate any assumption of a universe model for the calculation of the weighing factor (which is necessary when using absolute magnitudes and the source luminosity function).


\subsection{Normalised redshift distributions}

It is now straightforward to establish the normalised distribution $\ w_{Z_{d}}\left (z_{d}\right )$ of the deflector redshifts expected in the XXL field  
\begin{equation}
 w_{Z_{d}}\left (z_{d}\right ) = \frac{1}{\left <\tau\right >} \iint\! \left \{\frac{d \tau}{dz_{d}}\left (z_{s},z_{d},f_{X}\right )  d_{obs}\left (z_{s},f_{X}\right ) \right \} dz_{s} df_{X},
\label{eq_def_w_zd}
\end{equation} 
where the differential contribution $d \tau/dz_{d}$ of the redshift $z_{d}$ to a source lensing optical depth is given by Eq. \ref{eq_def_dtGL}.

Similarly, the normalised redshift distribution $w_{Z_{s}}\left (z_{s}\right )$ of the lensed sources is given by 
\begin{equation}
w_{Z_{s}}\left (z_{s}\right ) = \frac{1}{\left <\tau \right >} \int\! \tau\left (z_{s},f_{X}\right )   d_{obs}\left (z_{s},f_{X}\right ) df_{X}.
\label{eq_def_w_zs}
\end{equation}

\citet{Oguri2012} have derived by different means an expression for $ w_{Z_{d}}\left (z_{d}\right )$ and applied it to the sample of the SDSSQLS (see Eq. 23 in \citealt{Oguri2012}). This estimation was done by binning the redshift-magnitude plane and using the number of sources in the different bins as a weighing factor.
Their relation corresponds to the discrete equivalent of Eq. \ref{eq_def_w_zd}, where the integration runs over the redshift-magnitude plane, rather than the $\left (z_{s},f_{X}\right )$ one.

In \cite{Oguri2010}, the authors have derived a different expression for the redshift distribution of the lensed sources (Eq. 7 in \citealt{Oguri2010}), equivalent to Eq. \ref{eq_def_w_zs}, except that the integration runs over the absolute magnitude and the weighing function used in the integration corresponds to the expression of the joint probability density expressed in terms of the source luminosity function.

Finally, let us note that \cite{Mitchell2005} have also derived an estimation for the redshift distribution of the lensed sources in the CLASS survey, assuming the DNCF of the sources to be expressed as a single power-law expression, i.e. the amplification bias being thus constant for each source.
However, this assumption is very restrictive as QSOs usually show a DNCF that presents a break at a critical magnitude, and needs to be modelled by a double power law expression.

The use of $d_{obs}\left (z_{s},f_{X}\right )$ allows an easy calculation of $\left < \tau \right >$ through the survey, and the calculation of the expected normalised distributions $w_{Z_{s}}\left (z_{s}\right )$ and $w_{Z_{d}}\left (z_{d}\right )$, as a function of the redshift of the lensed sources and of the deflectors, respectively. These distributions are calculated without any assumption about the source population, as $d_{obs}\left (z_{s},f_{X}\right )$ may be directly estimated from the observed data, naturally including the observational biases.

In the next Section, we present the observational constraints used for the estimation of the joint probability density of the XXL survey.


\section{Observational constraints and $d_{obs}$ determination}
\label{section_obs_constraints}

%

\begin{figure*}
\centering
\includegraphics[height=0.8\columnwidth]{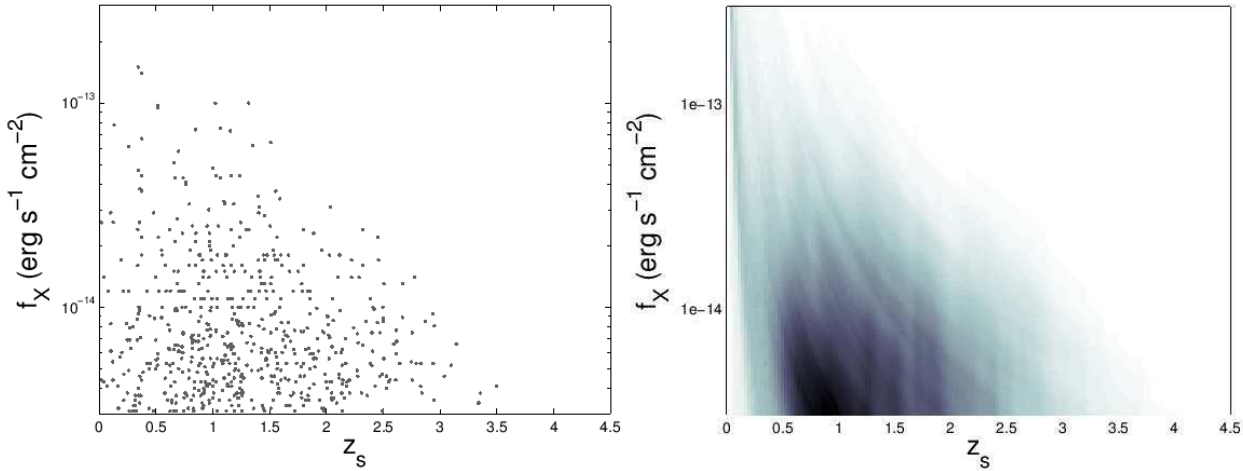}
\caption{The left panel displays the distribution in the $\left (z_{s},f_{X}\right )$ plane of the XMM-COSMOS source population \citep{Brusa2010}, restricted to $F_{\left [0.5-2\right ]keV} > 3 \times 10^{-15}$ erg cm$^{-2}$ s$^{-1}$ in the $\left [0.5-2\right ]$ keV band and with an optical counterpart brighter than $r=25$, for which a redshift estimate is available. This X-ray band flux limit corresponds to a detection probability of 0.5 in the XMM-LSS field, as defined in \citet{Elyiv2012}. The X-ray flux is shown on a logarithmic scale. The right panel displays the joint probability density function $d_{obs}\left (z_{s},f_{X}\right )$ in the $\left (z_{s},f_{X}\right )$ plane, corresponding to the XMM-COSMOS source population \citep{Brusa2010}, restricted to the X-ray band flux limit of the XMM-LSS fields.}  
\label{fig_Sx_z_distrib}
\end{figure*}
Ideally, the joint probability density of the XXL sources should be derived by doing a histogram of the XXL sources in the redshift-X-ray flux plane. However, the cross-correlation between the X-ray and (near-)optical data and the identification of the source type have not yet been performed. 
The XXL X-ray sources and their associated multi-wavelength data have characteristics similar to those of one of its sub-fields, the XMM-Large Scale Structure (XMM-LSS), covering 10.9 out of the 44.2 square degrees effectively covered by the XXL survey.
The XMM-LSS X-ray sources and their associated multi-wavelength data, presented in \citet{Chiappetti2013}, have optical counterparts taken from the CFHTLS W1 catalogue, down to the limiting magnitudes $i'\simeq 25$, $r'\simeq 25$, before correction for Galactic extinction
\footnote{After correction, the limiting magnitudes of the CFHTLS W1 catalogue are (expressed in AB magnitudes and considering a 5 sigma signal-to-noise ratio) $r \simeq 24.8$ and $i \simeq 24.5$.
See http://xxlmultiwave.pbworks.com/w/page/54613008/Optical and http://terapix.iap.fr/rubrique.php?id\_rubrique=268 for details.}.
The X-ray source classification and properties of the XMM-LSS field have already been determined (\citealt{Melnyk2013}).
The survey limiting flux in the soft band is $F_{\left [0.5-2\right ] keV} \simeq 3\times 10^{-15}$ erg s$^{-1}$ cm$^{-2}$ (with a detection probability of 0.5, as defined in \citealt{Elyiv2012}). Nevertheless, at the time of this work, the redshift estimate in the XMM-LSS sample is still ongoing.

Therefore, to retrieve the properties of the expected population to be detected within the XXL survey and their optical counterparts, we used data from a deeper field ($\sim 60$ ks exposure, compared to $\sim 10$ ks for XMM-LSS): the XMM-COSMOS field, to which we apply the flux cuts of the XXL in the X-ray and the optical, that we assume to be similar to those of the XMM-LSS. The XMM-COSMOS survey, covering a contiguous field of 2 square degrees is described in \cite{Brusa2010}, and has a limiting flux of $F_{\left [0.5-2\right ] keV} \simeq 5\times 10^{-16}$ erg s$^{-1}$ cm$^{-2}$ ($\simeq 1.4\times 10^{-15}$ considering the flux with with 50\% effective area coverage, see Fig. 6 in \citealt{Cappelluti2007}).
The flux limit at 50\% effective area coverage for the XMM-LSS and the COSMOS are in good agreement with the exposure time ratios as $3 \times 10^{-15}/1.4 \times 10^{-15} \sim \sqrt{60 ks / 10 ks}$.
Their optical catalogue with which the cross correlation was performed contains sources detected in at least one of the Subaru bands ($b,v,g,r,i,z$) down to an AB magnitude limit of $\sim 27$.
The COSMOS sample is almost complete down to $F_{\left [0.5-2\right ] keV} \simeq 3\times 10^{-15}$ erg s$^{-1}$ cm$^{-2}$ (X-ray detection probability of $\geq  0.98$, cfr \cite{Cappelluti2007}, Fig. 6) and 98\% of the X-ray sources have an optical counter part.

The joint probability density of the XXL survey is expected to be quite similar to that of the XMM-COSMOS field if we apply to this survey the same X-ray and optical flux cuts and take into account the different probability of detection for sources fainter than $2 \times 10^{-14}$ erg cm$^{-2}$ s$^{-1}$. Consequently, we first estimate the COSMOS joint probability density $d_{COSMOS}\left (z_{s},f_{X}\right )$ from a smoothed histogram of the sources from the XMM-COSMOS in the $\left (z_{s}, f_{X}\right )$ plane, where we apply the X-ray and $r$-band cutoffs of the XXL. We then take into account the XXL detection probability to determine $d_{obs}\left (z_{s},f_{X}\right )$.

In the left panel of Fig. \ref{fig_Sx_z_distrib}, we have represented the XMM-COSMOS sources in the $\left (z_{s},f_{X}\right )$ plane, with fluxes presented along a logarithmic scale.
We have only considered sources with a flux larger than $F_{\left [0.5-2\right ]keV} = 3 \times 10^{-15}$ erg cm$^{-2}$ s$^{-1}$ in the $\left [0.5-2\right ]$ keV band and with an optical counterpart brighter than $r= 25$ in the SDSS r-band. We have rejected all X-ray sources for which there was no redshift estimate. Whenever available, spectroscopic redshifts were preferred to photometric ones. 
The XMM-COSMOS catalogue presented in \citet{Brusa2010} contains 1797 sources in the 2 square degrees. After applying the X-ray and $r$-band cut-off (and excluding sources with no available $r$ magnitude), there are 630 sources of which 6 are excluded because of no available redshift. The final restricted COSMOS source sample contains 624 sources. 
The density of sources in the $\left (z_{s},f_{X}\right )$ plane is larger for the fainter fluxes, near redshift $z\sim 1$. At any given redshift, there are more sources with a lower flux. Finally, let us also point out the absence of sources with a high  flux at high redshift.

In order to estimate the joint probability density $d_{COSMOS}\left (z_{s},f_{X}\right )$ of the COSMOS sources, we have calculated a smoothed histogram of the COSMOS source distribution with the XMM cut of in the X-ray and $r$-band displayed on the left panel of Fig. \ref{fig_Sx_z_distrib}. 
For the convenience of the developed software, $d_{COSMOS}\left (z_{s},f_{X}\right )$ has been derived using a logarithmic scale for the X-ray fluxes. We have considered redshift intervals of 0.375 with bin centers separated by 0.0625 and logarithmic magnitude intervals of 0.3 separated by bins of 0.05. The derived COSMOS joint probability density is shown in the right panel of Fig. \ref{fig_Sx_z_distrib}. We have intentionally kept the same axis as in the left panel in order to clearly identify the similarities between the two figures. The grey scale indicates the values of $d_{COSMOS}\left (z_{s},f_{X}\right )$; the darker the grey, the higher the probability of finding a source.
The normalisation factor of $d_{COSMOS}\left (z_{s},f_{X}\right )$ is the number of sources (624) detected in the 2 square degrees of the COSMOS field, restricted to the XMM-LSS cutoffs.

To take into account the detection probability of the XXL survey, we multiply the $d_{COSMOS}\left (z_{s},f_{X}\right )$ in the right panel of Fig. \ref{fig_Sx_z_distrib} by the XMM-LSS detection probability as a function of the flux taken from \cite{Elyiv2012}, Fig. 10, the obtained distribution being $d_{obs}\left (z_{s},f_{X}\right )$. The normalisation factor of this distribution represents the number of sources (592.5) with $F_{\left [0.5-2\right ]keV} \ga 3 \times 10^{-15} $ erg cm$^{-2}$ s$^{-1}$ and $r<25$ that would be detected within the XXL survey in the 2 sq. deg. of the XMM-COSMOS field. From this normalisation factor, we may thus estimate the number of sources to be detected in the XMM-LSS and XXL-field with similar characteristics.
The expected numbers of sources in the different surveys are summarised in Table \ref{tab_GL_estimation}.
Assuming the final XXL catalogue to have similar characteristics as those of the XMM-LSS, this estimate of $d_{obs}\left (z_{s},f_{X}\right )$ is assumed to be valid for both the XMM-LSS and the XXL fields.

\begin{figure*}
\centering
\includegraphics[height=0.8\columnwidth]{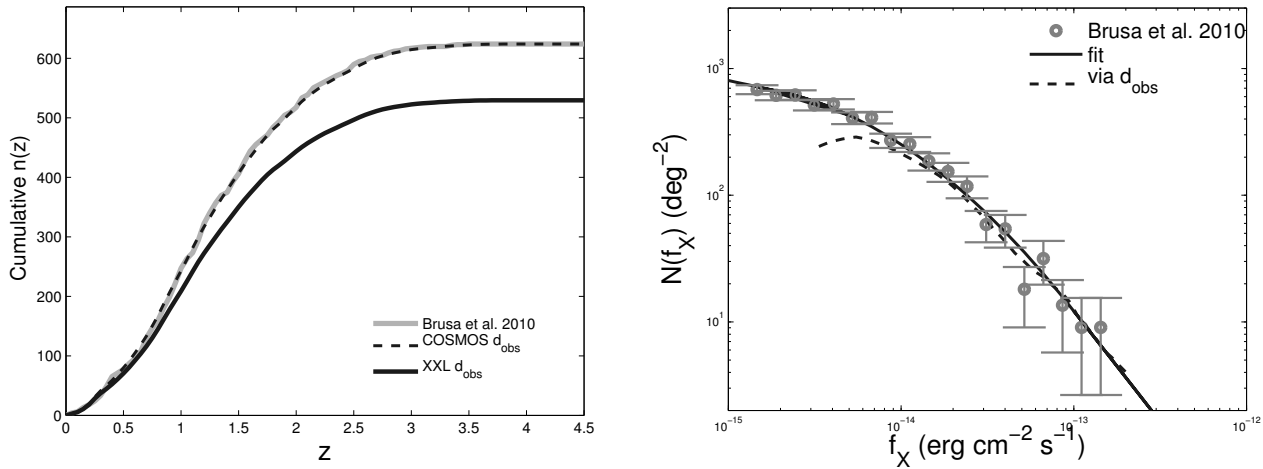}
\caption{Left panel : Cumulative function of the number density $N\left (z_{s}\right )$ of the sources as a function of the redshift for the XMM-COSMOS sources with $F_{\left [0.5-2\right ]keV} \ga 3 \times 10^{-15} $ erg cm$^{-2}$ s$^{-1}$ and $r<25$. We have also represented the cumulative redshift distribution inferred from the marginal distribution of the joint probability density function $d_{COSMOS}\left (z_{s},f_{X}\right )$ and $d_{obs}\left (z_{s},f_{X}\right )$, with and without considering the XXL detection probability detection (dashed and continuous black curves, respectively). 
Right panel: Differential number counts of the XMM-COSMOS sources as a function of the flux in the soft X-ray band, for all the COSMOS sources, which are used for the amplification bias calculation. We have only represented the X-ray flux range accessible to the XXL survey. We have also represented the marginal distribution obtained from the joint probability density function $d_{COSMOS}\left (z_{s},f_{X}\right )$, as well as the DNCF fit used for the calculation of the amplification bias.}  
\label{fig_source_z_histogram}
\end{figure*}

As a reliability test of the derived joint probability density $d_{obs}\left (z_{s},f_{X}\right )$, we verify its ability to represent the properties of the observed population of AGNs. 
In the left panel of Fig. \ref{fig_source_z_histogram}, we have represented as a continuous grey line the observed cumulative distribution as a function of the redshift of the XMM-COSMOS sources with an X-ray flux larger than $F_{\left [0.5-2\right ]keV} = 3 \times 10^{-15}$ erg cm$^{-2}$ s$^{-1}$ in the $\left [0.5-2\right ]$ keV band, with $r<25$ and a redshift estimation.

In the same figure, we have represented with a dashed dark-grey line the cumulative source redshift distribution derived from the COSMOS joint probability distribution when not considering the detection probability of the XXL survey. The cumulative redshift distribution is obtained by integrating $d_{COSMOS}\left (z_{s},f_{X}\right )$ over $f_{X}$ and by multiplying the number of AGNs detected per square degree in the COSMOS field, where we have applied the XXL cutoffs. 

The cumulative redshift distribution inferred from the distribution $d_{COSMOS}\left (z_{s},f_{X}\right )$ reproduces very well the observed cumulative source redshift distribution.

On the same figure, we have represented with a dark continuous line the cumulative redshift distribution derived from the final $d_{obs}\left (z_{s},f_{X}\right )$ when considering the detection probability of the XMM-LSS survey. We remark a very good agreement at low redshift between this distribution and the two previous ones, and a significant deviation above redshift $\sim 1$ due to the fact that the XXL survey does not detect all of the fainter sources, more numerous at higher redshifts.

In the right panel of Fig. \ref{fig_source_z_histogram}, we have represented the \textit{differential number counts function} (DNCF) of the XMM-COSMOS sources as a function of their X-ray soft band flux. The flux is shown along a logarithmic scale and the error bars are estimated considering a Poisson noise. We have here considered all the COSMOS sources brighter than $F_{\left [0.5-2\right ]keV} = 1 \times 10^{-15}$ erg cm$^{-2}$ s$^{-1}$. This DNCF is used to estimate $N_{f_{X}}\left (f_{X}\right )$ needed when calculating the amplification bias in Eq. \ref{eq_def_LCS}.
As previously mentioned, the calculation of the lensing optical depth necessitates the knowledge of the DNCF for fluxes fainter than the survey detection limit. 
We have therefore fitted the observed XMM-COSMOS DNCF $\log_{10}\left (n\left (f_{X}\right )\right )$ with a 3$^{rd}$ order polynomial in the flux range covered by the XMM-COSMOS and extrapolated the data linearly outside this flux range. The fit is shown as a continuous dark-grey line in the right panel of Fig \ref{fig_source_z_histogram}. This polynomial form of the DNCF is used for the calculation of the amplification bias in our simulations.

On the same figure, we have represented with a dashed line the DNCF per unit of solid angle derived from the joint probability density function $d_{obs}\left (z_{s},f_{X}\right )$, where we have restricted the COSMOS sources to the XXL cutoffs and taken into account the detection probability of the survey. This is obtained by integrating $d_{obs}\left (z_{s},f_{X}\right )$ over $z_{s}$ and by multiplying by the number $N_{AGN}/\Omega$ of AGNs per square degree. 
This curve is matching fairly well the fitted curve for the brighter sources and is lower for the fainter sources, where the XXL survey only detects part of the sources and where the fraction of excluded sources with $r>25$ is higher.

The small size of the source sample in the XMM-COSMOS field at low redshift (due to the small survey angular coverage) does not permit a better estimate of $d_{obs}\left (z_{s},f_{X}\right )$ at these low redshift values, because of a large scatter in the observed data at low redshift, especially for the brighter sources. 
Nevertheless, as these very bright and low redshift sources are very rare in the survey and as their lensing probability is very small (because of their very low redshift), this does not have a large impact on our simulations. 
These problems will be reduced in the XXL survey, for which the angular coverage is $\sim 20$ times larger than that of the XMM-COSMOS field. This confirms, as mentioned earlier, that the COSMOS sample is not large enough to determine the distribution $d_{obs}\left (z_{s},f_{X}, r\right )$ of the sources in the 3-D space $\left (z_{s},f_{X}, r\right )$.

\section{Results}
\label{section_results}


\subsection{Mean lensing optical depth}

Using the joint probability density $d_{obs}\left (z_{s},f_{X}\right )$ described in Section \ref{section_obs_constraints}, we have computed the mean lensing optical depth $\left < \tau\right >$ for the XXL survey, integrating numerically Eq. \ref{eq_def_mean_tGL}.
As $d_{obs}\left (z_{s},f_{X}\right )$ is assumed to be identical for the XXL and the XMM-LSS surveys, the results are valid for both surveys.

To calculate the lensing optical depth, we modelled the deflector population with SIS and SIE mass distributions (Eq. \ref{eq_t_SIS} and Eq. \ref{eq_t_SIE}). We have computed the average optical depth for different values of the minimum image separation $\theta_{mis}$ resolvable at optical wavelengths.
When calculating the cross section, we have considered the detection of the lensed images to be achievable down to an angular separation $\theta_{mis}$ (independently of their relative amplification). Although this is not strictly accurate, we have made this assumption for the following reason. The regions contributing the most to the lensing cross section are those where the source is located close to the caustic curves, as these are the most amplified and thus benefit the most from the amplification bias. For these configurations, the lensed images that are the closest to each other (and also the brightest) are those formed on each side of the tangential critical curve. These lensed images have a very similar amplification \citep{Kormann1994}. Consequently, for the detection of these lensed images, the critical parameter is the smallest angular distance under which these point-like images cannot be resolved, independently of their brightness. 
In practice however, for lensed images close to each other, the minimum angular distance at which the point-like images can be disentangled is dependent on the flux difference between the images, especially at very small angular distances. For the analysis of the final XXL sample, the angular selection function (characterizing the smallest angular distance detectable as a function of the relative amplification of the lensed images) will have to be determined precisely.

\begin{figure}
\centering
\includegraphics[width=\hsize]{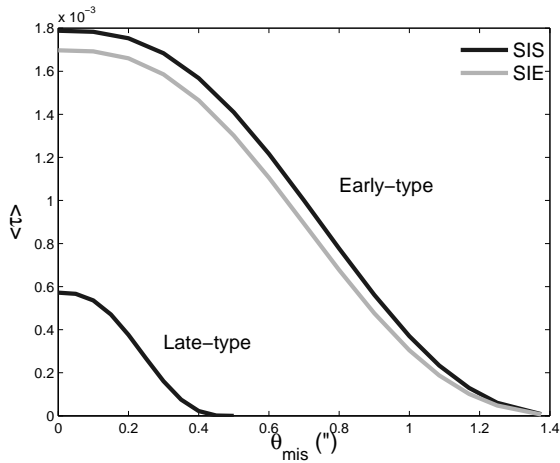}
\caption{Average lensing probability $\left < \tau \right >$ calculated for the XXL and XMM-LSS fields as a function of the minimum angular separation $\theta_{mis}$ resolvable by the survey in the optical domain, when modelling the deflector population with the SIS and SIE mass distributions. For comparison, we have also computed the case of the late-type galaxies, modelled by means of SIS mass distributions.}
\label{fig_t_vs_AR}
\end{figure}

Fig. \ref{fig_t_vs_AR} displays the behaviour of the average optical depth $\left < \tau\right >$ as a function of the minimal image separation $\theta_{mis}$ resolvable in the optical survey. We display the results when modelling the deflector population with either SIE or SIS mass profiles. In the former case, we have calculated the total probability of having a multiply-imaged source, independently of the number of the lensed images.

For a perfect instrument (i.e. $\theta_{mis}=0"$), when considering deflectors modelled with SIE mass distributions, we find an average lensing probability $\left <\tau_{SIE}\right > = 1.698 \times 10^{-3}$. When modelling the deflectors with an SIS mass distribution, the mean lensing probability is $\left <\tau_{SIS}\right > = 1.788 \times 10^{-3}$.  Modelling the deflectors with SIS mass distributions thus leads to a slightly larger average lensing probability by $\sim 5\%$. 
The reason for this is that the SIE mass distribution was introduced in order to preserve the projected mass inside a same area but does not conserve the geometrical cross section nor the amplification probability distribution. In other words, the area inside the caustic curve of an SIS deflector is always larger or equal to that included inside the caustic curves of an SIE deflector. 
When averaging the SIE lensing cross section over the deflector ellipticity distribution, this leads to an equivalent SIE lensing cross section smaller than that of the SIS case.
Furthermore, for these two types of deflectors, the probability of producing a multiply-imaged source with a given total amplification slightly differs. The impact of the amplification bias is thus different for the two deflector models. Consequently, the slight differences between $\left <\tau_{SIS}\right >$ and $\left <\tau_{SIE}\right >$ depend on the DNCF as a function of $f_{X}$ for the source population, which varies from one survey to another. The overestimate of $\left <\tau_{SIS}\right >$ thus has to be estimated independently for each survey.
This boost of the average lensing optical depth in the SIS model was studied by \cite{Huterer2005} who concluded that the ellipticity in the deflector mass distribution decreases the mean lensing optical depth. A boost of the SIS model by a few percent may be expected in a survey with sources showing a steep DNCF. 

Both $\left <\tau_{SIE}\right >$ and $\left <\tau_{SIS}\right >$ decrease for increasing values of the parameter $\theta_{mis}$. When considering a finite resolution of the instrument, some of the lensed images formed are angularly too close to each other and are detected as a single point-like object. Consequently, the probability of \textit{detecting} the lensed images decreases as $\theta_{mis}$ increases.

\begin{table*}
\caption{Estimate of (1) the number of sources with an X-ray flux greater than $3\times 10^{-15}$ erg cm$^{-2}$ s$^{-1}$ in the $\left [0.5-2\right ]$keV range and $r<25$ and (2) the number of detected multiply-imaged sources in the XMM-COSMOS, XMM-LSS and XXL surveys. The estimates of the number of sources are extrapolated from the XMM-COSMOS catalogue taking into account the angular coverage of the different surveys. For the estimate of the number of multiply-imaged sources, the numbers in parentheses correspond to the SIS case.}             
\label{tab_GL_estimation}      
\centering          
\begin{tabular}{l | c c c c c c }     
\hline\hline       
Survey & Coverage & Number of sources & Lensed sources    & Lensed sources     & $>$2 images  & $>$2 images\\ 
	   &(deg$^{2}$)&                   &($\theta_{mis}=0"$)&($\theta_{mis} =0.45"$)& ($\theta_{mis} = 0"$) &($\theta_{mis} =0.45"$) \\ 
	   &&                   &SIE (SIS)&SIE (SIS)&  & \\ 	   
\hline                    
\\
XMM-COSMOS & \textbf{2} & \textbf{621}  & - & - & - & - \\  
\\
XMM-COSMOS & \textbf{2} & \textbf{529.5}  & - & - & - & - \\  
(with XXL $P_{detection}$)& &  &  &  &  \\  
\\
XMM-LSS & \textbf{10.9} & 2885    & 5 (5) & 4 (4) & 0 & 0 \\
\\
XXL        & \textbf{44.2} & 11701   & 20 (21) & 16 (17) & 1 & 1 \\
\\
\hline
\hline
       &   &   & $\left < \tau\right >$ & $\left < \tau\right >$  & $\left < \tau_{>2}\right >/\left < \tau\right >$ & $\left < \tau_{>2}\right >/\left < \tau\right >$\\
       &   &   & ($\theta_{mis}=0"$) & ($\theta_{mis}=0.45"$)  & ($\theta_{mis}=0"$) & ($\theta_{mis}=0.45"$)\\
\hline
\\
SIE&   &   & $1.698\times 10^{-3}$ & $1.384 \times 10^{-3}$ & 0.0718 & 0.0564 \\
\\
SIS&   &   & $1.788 \times 10^{-3}$ & $1.489 \times 10^{-3}$ & - & - \\
\\
SIS (Late)&   &   & $5.719 \times 10^{-4}$ & $1.629 \times 10^{-6}$ & - & - \\
\hline

\hline                  
\end{tabular}
\end{table*}

In the present case, the multiple lensed images will be searched for among the optical counterparts of the point-like X-ray sources, because of the better angular resolution in the optical domain. The ground-based observations are limited by the atmospheric seeing. In the northern XXL-fields, the CFHT in the $r$-band have a typical seeing of $0.7$" (\citealt{Salmon2009}) and for the southern hemisphere the typical seeing with the Blanco telescope is $\sim 0.9"$ (\citealt{Desai2012}). We thus considered our full width at half maximum of the PSF to be homogeneous over the entire sample and equal to $\sim 0.9"$.

Thanks to point spread function fitting techniques, we can hope to resolve multiple point-like lensed images down to half the full width at half maximum for lensed images with a same amplification, which constitute the configurations contributing the most to the lensing cross section as previously explained. Consequently, the typical $\theta_{mis}$ value achievable is expected to be $\theta_{mis} \sim 0.45"$ (in practice, as stressed previously, $\theta_{mis}$ depends of the relative amplification of the lensed images, which will have to be taken into account when analyzing the final XXL sample).
For this value we have $\left <\tau_{SIE}\right > = 1.384 \times 10^{-3}$ and $\left <\tau_{SIS}\right > = 1.489 \times 10^{-3}$. The slight overestimate of the SIS mean lensing value relatively to that of the SIE model thus increases with $\theta_{mis}$ and reaches $\sim 10\%$ for $\theta_{mis} = 0.45"$.

For comparison, we have computed the evolution of the mean optical depth as a function of $\theta_{mis}$ for the population of late-type galaxies modelled by means of an SIS mass distribution. As for the case of the early-type galaxies, we have considered the comoving density of late-type galaxies to be constant with the redshift and we have used the VDF parameters determined by  \citet{Chae2010} in the local universe, i.e. $$\left [\Phi_{\ast},\sigma_{\ast},\alpha, \beta \right ] = \left [66\times 10^{-3} \,\mathrm{h}^{3}\, \mathrm{Mpc}^{-3}, 91.5 \,\mathrm{km}\, \mathrm{s}^{-1}, 0.69, 2.10 \right ].$$
The results are displayed in Fig. \ref{fig_t_vs_AR}. When considering a perfect instrument ($\theta_{mis}=0$"), the average lensing optical depth associated to the late-type galaxies is about a third of that of the early-type ones (i.e. $\left <\tau\right > = 5.719 \times 10^{-4}$). Nevertheless, the decrease of $\left <\tau\right >$ with $\theta_{mis}$ is steeper than in the early-type galaxy case. Indeed, late-type galaxies are less massive and lead to smaller typical angular separations of the multiple lensed images, which are not disentangled in the seeing limited images. For $\theta_{mis}= 0.45$", the average lensing optical depth due to late-type galaxies is found to be $\left <\tau\right > = 1.629 \times 10^{-6}$, three orders of magnitude lower than that of the early-type galaxies. The expected contribution of the late-type galaxies in our sample of gravitationally lensed sources is thus negligible, which validates our assumption of only considering the population of early-type galaxies as the deflectors for the XXL lensed source sample.


Our estimation of the contribution of late-type galaxies to the lensed sources is surprizingly low compared to the observed fraction of late-type lenses in existing samples. Furthermore, it suggests that late-type lenses with image separations larger than 0.5" are extremely rare, which is also in contradiction with observed samples.
Indeed, 2 out of the 13 CLASS lenses from the statistical sample are likely to be produced by late-type galaxies (B0218 and B1933, see \citealt{Browne2003})
and out of the 26 lensed QSOs of the SDSSQLS statistical sample, one is possibly due to a late-type galaxy (J1313) and has an angular separation larger than 1". 
We do not fully understand the reason for these discrepancies. A possible cause of error is the effect of the lens environment, not considered in this work, which may lead to an additional gravitational shear. Huterer et al. 2005 have shown that the external shear broadens the distribution of angular separation between the lensed-images without changing its average value. It may therefore increase the fraction of events with an angular separation larger that 1". In the CLASS and SDSSQLS statistical samples, all the lenses produced by late-type galaxies with and angular separation larger than 1" required external shear to accurately model the position and relative amplification of the lensed images (see \citealt{Suyu2012}, \citealt{Sluse2012} and references therein). 
Another possible source of discrepancy is the VDF used for late-type galaxies (from \citealt{Chae2010}) which does not come from direct measurements (as in the case of the early type VDF, \citealt{Choi2007}). It is inferred from the local luminosity function of late-type galaxies using the Tully-Fisher relationship (taking into account its dispersion) and assuming a conversion between the circular velocity $v_{c}$ and $\sigma$ to be that of an SIS profile (i.e. $\sigma = v_{c}/\sqrt{2}$). \cite{Chae2010} stresses that lensing statistics of late-type galaxies might necessitate to consider the circular velocity function (rather than the VDF) and consider more realistic mass distribution models. 
Furthermore, type-specific LF (and VDF) are still potentially biased by misclassification of the galaxy types and, according to \cite{Park2005}, the work of \cite{Chae2010} has a $\sim 10$\% classification mismatch. 
We therefore advise to use with caution the results concerning the late-type population statistics.


Table \ref{tab_GL_estimation} summarises the number of multiply-imaged sources expected in the different surveys as well as the expected number of detected events.
Assuming the sources detected in the XXL survey will have the same properties as those in the XMM-COSMOS to which we applied the same flux limits in the X-ray and optical bands (except when estimating the amplification bias) and when accounting for the detection probability of the XXL, we expect the detection of 11701 sources with $F_{\left [0.5-2\right ]keV}> 3 \times 10^{-15}$ erg cm$^{-2}$ s$^{-1}$ and $r<25$ in the 44.2 sq.deg. of the survey. Among these sources, we expect 21 (20 in the SIE case) to be multiply imaged, out of which 17 (16 in the SIE case) should be detected assuming $\theta_{mis} = 0.45"$.

\begin{figure}
\centering
\includegraphics[width=\hsize]{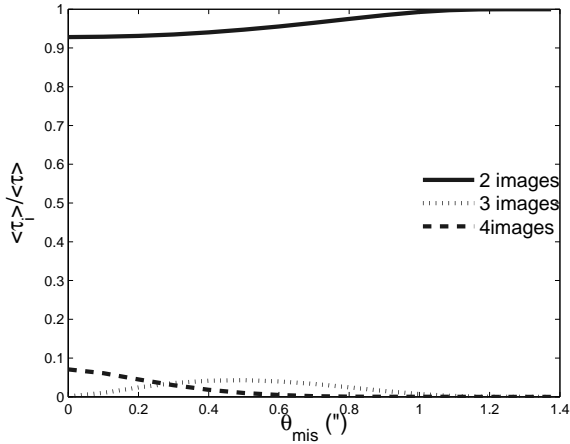}
\caption{Fraction $\left <\tau_{SIE,i}\right > /\left <\tau_{SIE}\right > $ of the lensing events with a given number $i=2,3\, \mathrm{or}\, 4$ of images as a function of the survey minimum angular separation $\theta_{mis}$, when modelling the deflector population with the SIE mass distribution. 3-image configurations are quads with 2 blended point-like images.}
\label{fig_trel_vs_AR}
\end{figure}

When calculating the lensing probability of a source to be lensed through Eq. \ref{eq_t_SIE}, considering a population of deflectors modelled with SIE profiles, we may also calculate the probability $\tau_{SIE,i}$ of a source to be lensed with the formation of a given number $i$ of lensed images.
We have calculated the average value of $\left <\tau_{SIE,i}\right >$ for the population of sources to be detected within the XXL and XMM-LSS fields using Eq. \ref{eq_def_mean_tGL}, for different values of the $\theta_{mis}$ parameter. The results are displayed in Fig. \ref{fig_trel_vs_AR} where we have plotted as a function of the value of the $\theta_{mis}$ parameter, the fraction of lensing events composed of $i$ lensed images $\left <\tau_{SIE,i}\right > /\left <\tau_{SIE}\right > $ relatively to the total average lensing probability $\left <\tau_{SIE}\right > $.

The fraction of lensing events with the formation of two images is always the highest. This is a consequence of the scarcity of very elliptical deflectors (see \citealt{Choi2007}, Fig. 13 for the axis ratio distribution of early-type galaxies).
In the case of a perfect instrument (i.e. $\theta_{mis}=0"$ ), we find that 93\% of the lensed sources are composed of 2 images and this fraction increases with the value of the $\theta_{mis}$ parameter.

For a perfect instrument, the lensed sources with formation of more than 2 images is composed of quads (i.e. 4-image configurations). As $\theta_{mis}$ increases, some of the 4-lensed image configurations, due to their too small angular separation, have only 3 point-like images detected (quads with 2 blended point-like images).
Out of the 20 lensed sources formed in the XXL population, only one is expected to be detected with more than 2 images.

The fraction of lensing systems with more than 2 images is roughly consistent with the results of \cite{Oguri2010} who have calculated the expected number of gravitationally lensed quasars in wide-field optical surveys. For a survey with a limiting magnitude $i = 25$, these authors find a fraction of a little more than $\sim 10$\% of quads (formation of 4 images) against $\sim 7$\% in our simulations. The slight difference is most likely due to the different ellipticity distribution of the deflectors considered (\citealt{Oguri2010} considers a combination of oblate and prolate three dimensional deflectors with a Gaussian distribution of their ellipticity) as well as their consideration of an additional external shear due to the lens environment.
\cite{Huterer2005} showed that the external shear increases the fraction of quads in a sample of lensed sources.


\subsection{Redshift distributions}

\begin{figure}
\centering
\includegraphics[width=\hsize]{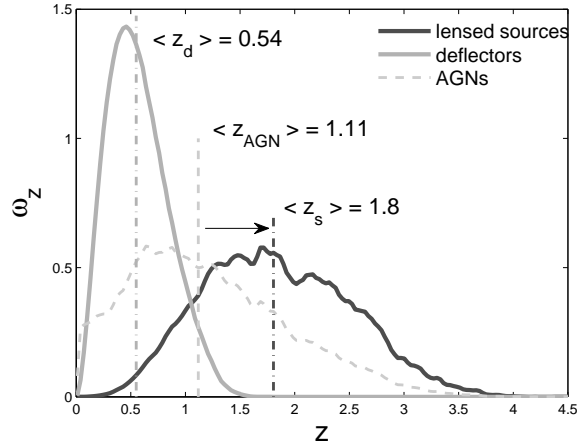}
\caption{Normalised redshift distributions of the deflectors, the lensed sources and all AGNs. For each distribution, we have indicated the median redshift value.}
\label{fig_dist_vs_z}
\end{figure}

Using the joint probability density $d_{obs}\left (z_{s},f_{X}\right )$ described in Section \ref{section_obs_constraints}, we have computed the mathematical expectation of the normalised redshift distribution of the deflectors $w_{Z_{d}}\left (z_{d}\right )$ and of the lensed sources $w_{Z_{s}}\left (z_{s}\right )$, numerically integrating Eqs. \ref{eq_def_w_zd} and \ref{eq_def_w_zs}, respectively.
We have considered $\theta_{mis}=0.45$" and we have modelled the deflectors with SIS mass distributions. The normalised redshift distributions $w_{Z_{d}}\left (z_{d}\right )$ and $w_{Z_{s}}\left (z_{s}\right )$ are shown as a function of the redshift in Fig. \ref{fig_dist_vs_z}. For comparison, we have represented the marginal distribution as a function of the redshift of the joint probability density, i.e. the normalised distribution as a function of the redshift of all the sources (independently of the fact that they are being lensed).

On this figure, we have also illustrated the median value of the observed redshift for the different distributions. The redshift distribution of the lensed sources is shifted towards a higher redshift compared to that for all the sources. As the source redshift increases, so does its geometrical lensing volume (i.e. the volume in which the presence of a deflector leads to the formation of multiple lensed images). Consequently, sources  with a higher redshift tend to have a higher lensing probability. The mathematical expectation for the redshift moves from $\left <z_{AGN}\right > \simeq 1.11$ for the entire source population to $\left <z_{s}\right > \simeq 1.8$ for the lensed sources.

For redshifts larger than $\left <z_{AGN}\right >$, the distribution of the lensed sources does not appear as a smooth function of the redshift, and we clearly see the presence of bumps or redshift ranges with a probability excess compared to a smooth decreasing function of the redshift. These probability over-densities of lensed source detections correspond to redshift ranges in which strong emission lines of the AGNs enter in the optical SDSS r-band, in which the optical counterparts are searched for; they are thus a consequence of a selection bias. The presence of the emission lines in the r-band increases their probability of being detected, compared to that of a source with only a continuum-like spectrum. 
For example, the MgII line at 279.8 nm enters the r-band filter in the redshift range 2 - 2.42, for which we see an over-density in the lensed source redshift distribution.

The redshift distribution of the lensed sources corresponds to the probability density from which the detected lensed source redshift may be considered as a random event, thanks to the fact that, in this case all the lensed sources are detected, and may have their redshift estimated. Given a large enough sample of lensed sources, this distribution could be retrieved from a normalised histogram of the lensed sources, as a function of their redshift.

In Fig. \ref{fig_dist_vs_z}, we have also represented the normalised distribution $\omega_{Z_{d}}$, as a function of the redshift, for the deflectors involved in the formation of multiply-imaged sources. The deflector redshift median value is $\left <z_{d}\right > \sim 0.54$ and the most probable value is below $z\la 0.5$. The contribution of deflectors with $z\ga 1$ is very small. This comforts our assumption of a non-evolving deflector population in the calculation of the lensing optical depth, as the population involved is mainly located at low redshift.

In the case of the deflector distribution, the observed redshift distribution of the deflectors involved in the formation of multiple images of a source will be highly biased as most of them are not bright enough to be detected. 

\cite{Oguri2010} have also estimated the expected lens redshift distribution for lensed quasars detected in optical imaging surveys. The deflector redshift distributions are marginally consistent, although the distribution found in this work peaks at lower redshifts ($z_{max}\sim 0.6$ for \citealt{Oguri2010} and $z_{max}\sim 0.5$ in the present work). 
The expected source redshift distribution is also shifted towards lower redshift in our study.
This difference in the lensed source and deflector redshift distributions comes from the difference in the source distribution: in the present work, the rather bright X-ray flux cut-off tends to reject sources at high redshift, that are included in the sample of \cite{Oguri2010}.


\subsection{Influence of the cosmological model}

\begin{figure}
\centering
\includegraphics[width=\hsize]{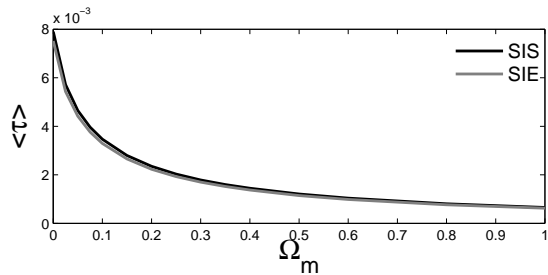}
\caption{Average lensing probability as a function of the cosmological matter density parameter $\Omega_{m}$.}
\label{fig_t_vs_Om}
\end{figure}

The probability for a source with a known apparent flux and redshift to be gravitationally lensed with the formation of multiple images depends on the cosmological model. This may be seen for instance through the dependence of the infinitesimal light-distance element $cdt/dz$ on the cosmological mass density parameter $\Omega_{m}$ in Eq. \ref{eq_def_cdtdz}, as well as through the dependence of the lensing cross section on $\Omega_{m}$ (see Eqs. \ref{eq_def_LCS} and  \ref{eq_def_b0}) via the definition of the angular diameter distances.
If we consider a flat expanding FLRW universe model, the only dependence of the lensing probability on the universe model is made through the cosmological mass density parameter $\Omega_{m}$. 
Indeed, although a flat expanding FLRW is totally characterised by $\Omega_{m}$ and $H_{0}$, an increase in the value of $H_{0}$ will only act as a scaling factor (decreasing the lensing volume while increasing the density of deflectors).
So far, we have considered a FLRW flat universe with $\Omega_{m}=0.3$ and $H_{0} = 70$ km s$^{-1}$ Mpc$^{-1}$. In Fig. \ref{fig_t_vs_Om}, we display the behaviour of the average lensing probability $\left <\tau\right >$ for the XXL sources as a function of $\Omega_{m}$, for a flat universe with $H_{0} = 70$ km s$^{-1}$ Mpc$^{-1}$.  
We have computed both $\left <\tau_{SIE}\right >$ and $\left <\tau_{SIS}\right >$. Both models lead to the same behaviour of the mean lensing probability as a function of $\Omega_{m}$ and, here as well, we observe that the SIS mass distribution leads to a slight overestimate of the average lensing probability when compared to that corresponding to the SIE model, whatever the value of $\Omega_{m}$.

There is a very strong dependence of the expected fraction of lensed sources in the survey on $\Omega_{m}$. For this reason, the statistics of gravitational lensing in a well-defined sample of sources has been widely used to probe the value of $\Omega_{m}$ and test dark energy models (\citealt{Turner1984}, \citealt{Fukugita1990}, \citealt{Turner1990}, \citealt{Surdej1993}, \citealt{Keeton1998b}, \citealt{Chae2002}, \citealt{Keeton2002}, \citealt{Ofek2003}, \citealt{Mitchell2005}, \citealt{Cao2012}). 

The XXL sample on its own will not allow better constraints on the value of $\Omega_{m}$ than recent lens surveys such as the SDSS-LQS. \cite{Oguri2012} for instance constrained $\Omega_{Lambda}$ to $\Omega_{Lambda} = 0.79^{+0.06 (\mathrm{stat.})}_{-0.07}\pm 0.06(\mathrm{syst.})$ on the basis of 19 lenses from the SDSS-QLS statistical sample. The XXL lensed source sample will thus be combined with other recent surveys (including the SDSS-QLS) to better constrain the cosmological parameters.

\section{Considering the $r$-band cut-off}
\label{section_rbandCutoff}
   
\begin{figure*}
\centering
\includegraphics[height=0.8\columnwidth]{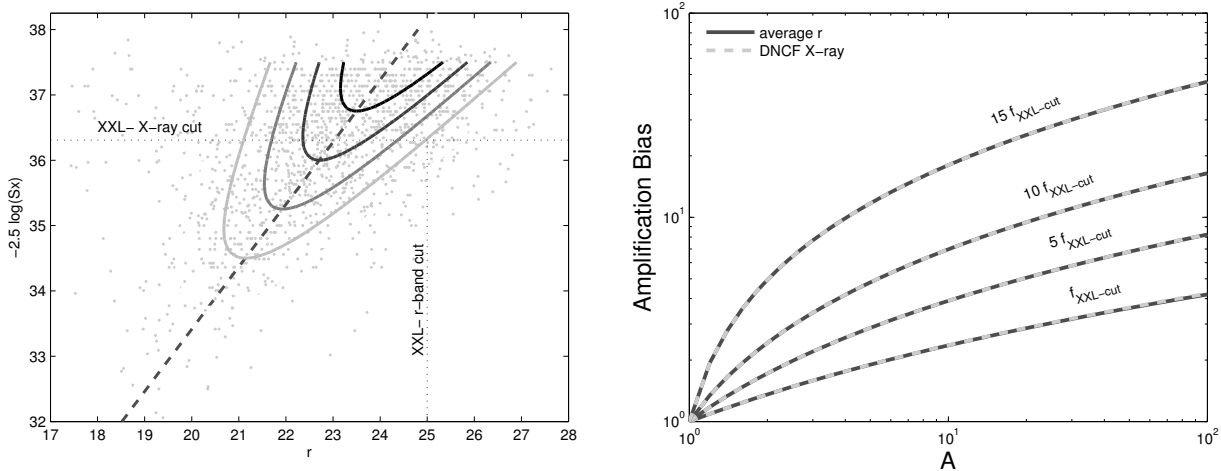}
\caption{\textit{Left :} distribution of the COSMOS sources in the $\left (f_{X},r\right )$ plane. We have represented the X-ray fluxes in terms of the X-ray magnitudes $m_{x}=-2.5\log f_{X}$, and have represented the X-ray and r-band cut-off of the XXL survey. We have also represented the contour plot of the COSMOS source density in the $\left (f_{X},r\right )$ plane. \textit{Right :} amplification bias as a function of the total amplification of the lensing event when considering only the DNCF as a function the X-ray flux $f_{X}$ (continuous dark grey line) and when averaging over the $r$-band magnitude distribution of the sources. }
\label{fig_bias_vs_mx}
\end{figure*}
In the previous chapter, when calculating the amplification bias in Eq. \ref{eq_def_LCS}, we have only considered the distribution of the sources as a function of their flux $f_{X}$ in the X-ray band and we have considered the lensing optical depth $\tau\left (z_{s}, f_{X}\right )$ to be only a function of the source redshift and its flux in the X-ray band.
Ideally, the source $r$-magnitude in the optical domain should also be considered for the calculation of $\tau\left (z_{s}, f_{X},r\right )$ and $d_{obs}\left (z_{s}, f_{X},r\right )$ and the calculation of the mean lensing optical depth $\left <\tau\right >$ in Eq. \ref{eq_def_mean_tGL} should include the integretation over $r$.

The joint probability density $d_{obs}\left (z_{s}, f_{X},r\right )$ may be decomposed as 
\begin{equation}
d_{obs}\left (z_{s}, f_{X},r\right ) = d_{obs}\left (z_{s}, f_{X}\right ) d\left (r|z_{s}, f_{X}\right )
\end{equation}
where $d\left (r|z_{s}, f_{X}\right )$ is the normalised distribution as a function of $r$ for sources with a redshift and an X-ray flux in the ranges $\left [z_{s}, z_{s}+dz_{s}\right ]$ and $\left [f_{X}, f_{X} +df_{X}\right ]$, respectively, and Eq. \ref{eq_def_mean_tGL} remains correct if we consider the optical depth $\tau_{\left <r\right >}\left (z_{s},f_{X}\right )$ averaged over the $r$ magnitudes
\begin{equation}
\tau_{\left <r\right >}\left (z_{s},f_{X}\right ) = \int \tau\left (z_{s},f_{X},r\right ) d\left (r|z_{s}, f_{X}\right ) dr.
\end{equation}
Equivalently, $\tau_{\left <r\right >}\left (z_{s},f_{X}\right )$ can be calculated through Eq. \ref{eq_def_tGL} if we consider an average lensing cross section $\Sigma_{\left <r\right >}$
\begin{equation}
\Sigma_{\left <r\right >} = b_{0}^{2} \iint\limits_{S_{y}} B_{\left <r\right >} d\mathbf{y},
\end{equation}  
where we have defined the amplification bias $B_{\left <r\right >}$ averaged over the $r$-magnitudes
\begin{equation}
B_{\left <r\right >}=\int \dfrac{N\left (z_{s},f_{X}/A_{X}, r + 2.5 \log\left (A_{r}\right )  \right )}{N\left (z_{s},f_{X}, r\right )} d\left (r|z_{s}, f_{X}\right ) dr,
\label{eq_bias_meanr}
\end{equation}
where $N\left (z_{s},f_{X}, r\right )$ represents the density of sources with a redshit $z_{s}$, having an observed flux $f_{X}$ in the X-ray and a $r$-band magnitude $r$, $A_{X}$ and $A_{r}$ are the amplifications due to the lensing event in the X-ray and $r$-band, respectively. 
If the source is point-like, we may then assume that $A_{X}=A_{r}$. If not, the ratio $A_{X}/A_{r}$ will depend on the size of regions in the AGN emitting the X-ray and the optical fluxes, respectively, as well as the positions of these regions with respect to the caustic curves. The expectation of the ratio $A_{X}/A_{r}$ will thus necessitate heavy simulations to be performed. As a first approximation we consider point-like sources.

In this work, the amplification bias $B_{X-ray}$ has been calculated so far uniquely on the basis of the X-ray flux through the relation 
\begin{equation}
B_{X-ray} = \dfrac{N\left (f_{X}/A_{X} \right )}{N\left (f_{X}\right )},
\label{eq_bias_Xray}
\end{equation}
where we have neglected the redshift dependence of the sources as a function of their redshift because of a too small source sample to correctly characterise this possible redshift dependence.
In the following we will compare the amplification bias obtained through Eqs. \ref{eq_bias_meanr} and \ref{eq_bias_Xray}.

To calculate rigorously the amplification bias through Eq. \ref{eq_bias_meanr} we thus need to determine the distribution of the source population in the $\left (z_{s},f_{X},r\right )$ space. Our observational data from the COSMOS sample is not large enough to estimate this distribution over the whole area probed by the survey. Consequently, we assume the distributions $N\left (z_{s},f_{X}, r\right )$ and $d\left (r|z_{s},f_{X}\right )$ to be independent of the redshift. 

Let us now estimate the distribution $N\left (f_{X},r\right )$ on the basis of the distribution of the COSMOS sample in the $\left (f_{X},r\right )$ plane, represented on the left panel of Fig. \ref{fig_bias_vs_mx}. We have represented all the COSMOS sources for which the soft X-ray flux and $r$-band magnitude is given in the sample of \cite{Brusa2010}. As 98\% of the detected X-ray sources have an optical counterpart in the r-band for X-ray sources brighter than the XXL 0.5 detection probability cut, we can consider this sample to be a complete sample of X-ray sources brighter than the X-ray cut, for which we know the optical counterpart.

To estimate $N\left (f_{X},r\right )$ we have proceeded as follows. We bin in terms of the $m_{x} = -2.5\log f_{X}$ magnitude, with a bin width of $dm_{x}=0.5$ and bin centres ranging from 33.5 to 37. 
In each $m_{x}$ bin, we construct the histogram of the sources as a function of the $r$-band magnitude, using bins with a width of $dr = 0.5$ and bin centres ranging from 17 to 28 with step of 0.5. Each histogram as a function of $r$ is fitted by means of a Gaussian profile. The number of sources in each bin is then divided by $dm_{x}dr$.

We thus obtain the evolution of the Gaussian fit parameters (i.e. the amplitude $A_{G}$, the average position $r_{G}$ and the standard deviation $\sigma_{G}$) as a function of the $m_{x}$ bin and we fit the dependence of these parameters as a function of $m_{x}$ by a linear law in the $m_{x}$ range 33.5 to 37. We model the density of the sources in the $\left (f_{x},r\right )$ plane by 
\begin{equation}
N\left (f_{X},r\right ) = A_{G}\left (m_{x}\right ) \exp\left (-\left (\dfrac{r-r_{G}\left (m_{x}\right )}{2\sigma_{G}\left (m_{x}\right )}\right )^{2}\right ),
\label{eq_dobs_fx_r_model}
\end{equation}
The COSMOS data do not have enough bright sources to constrain $N\left (f_{X},r\right )$ for $m_{x}$ brighter than 33.5.
We have represented the iso-density contours of the calculated function $N\left (f_{X},r\right )$ on the left panel of Fig. \ref{fig_bias_vs_mx}, showing a very good agreement between the observed distribution of the COSMOS sources and the modelled density function.

We now use the modelled function $N\left (f_{X},r\right )$ to calculate the amplification bias in Eq. \ref{eq_bias_meanr}, for sources with different X-ray to optical flux ratios. Assuming the amplification induced by the lensing event in the X-ray and the optical are identical, when amplified, a source is displaced in the $\left (m_{x},r\right )$ parallel to the dashed line shown on the left panel of Fig. \ref{fig_bias_vs_mx}. Consequently, the amplification bias is calculated thanks to the evolution of $N\left (f_{X},r\right )$ along this trajectory.

On the right panel, we represent the behaviour of the amplification bias as a function of the total amplification of the lensing event, for 4 sources with apparent X-ray fluxes equal to the XXL X-ray limiting flux $f_{XXL-cut}$, and 5, 10 and 15 times the value of the $f_{XXL-cut}$. The dashed light grey curve represents the amplification bias $B_{X-ray}$ obtained when only considering the DNCF of the XXL-sources as a function the X-ray flux, calculated through Eq. \ref{eq_bias_Xray}.
The continuous black curve corresponds to the average bias $B_{\left <r\right >}$ calculated thanks to Eq. \ref{eq_bias_meanr}. 
We see a perfect agreement between these curves and the amplification bias calculated only considering the DNCF as a function of the X-ray flux $f_{X}$. Consequently, the amplification bias calculated through Eq. \ref{eq_def_LCS} perfectly corresponds to the amplification bias averaged over the $r$-band magnitudes.

Thus, assuming a point-like source, we may calculate the amplification bias in the combined X-ray/optical data by considering uniquely the X-ray distribution of a complete and deeper sample, which validates the method introduced in Section \ref{section_theory}. Nevertheless, in the analysis of the final XXL sample, we will have to also consider the redshift dependence of the amplification bias, which will be made possible thanks to the much larger size of the sample.

\section{Conclusions}

We have calculated the expected statistical properties of the multiply imaged sources to be detected among the optical counterparts of the XXL point-like sources, modelling the deflectors successively with SIS and SIE mass profiles. We find that
\begin{itemize}
\item one expects the formation of 20 (21 using the SIS model) multiply-imaged AGNs out of which 16 (17 for the SIS case) should be detected among the optical counterparts with an angular resolution of $0.45$", and we only expect the detection of one gravitational lens system composed of more than 2 lensed images;
\item the expectations are consistent when modelling deflectors with SIE and SIS mass distributions, although the SIS model leads to a slight overestimate of the mean lensing probability. This overestimate is a function of the amplification bias and is thus different for each survey;
\item the late-type galaxy population should not contribute to the lensed sources to be detected;
\item although the detection is done simultaneously in the X-ray and in the optical domain, the amplification bias may be estimated from the X-ray flux distribution, as long as we consider a complete X-ray sample from a deeper survey and for point-like sources.
\end{itemize}

In this work, we have considered isothermal profiles to model the deflectors. This has allowed us to get first good estimates of the expected number of lensed AGN in the XXL survey. However, more detailed calculations ought to be carried out. Indeed, \cite{Auger2010} and \cite{Koopmans2009}, through the analysis of massive early-type deflectors from the SLACS survey, have found a slight deviation from the isothermal profile, with a steeper slope parameter ($\left <\gamma\right > = 2.078$ and 2.085, respectively, where the mass distribution evolves as $r^{-\gamma}$). If we consider two mass distributions with a same total mass, a steeper profile would lead to a higher fraction of the lens in the center, which will increase the Einstein angular radius, therefore leading to an increase in the lensing cross section of the deflector. This increase in the lensing cross section with the steepness of the radial mass profile has been put in evidence in a series of papers (\citealt{Mandelbaum2009} and \citealt{vandeVen2009}), where the authors studied the impact of galaxy shape and density profile on the selection biases in surveys for the detection of strong lenses. Because isothermal profiles are singular the authors analysed more realistic profiles in order to define the total mass.

Furthermore, \cite{Sonnenfeld2013} have shown from the study of the SL2S Galaxy-scale Lens Sample (with deflectors in the range $0.2<z<0.8$) that the mass density profile of early-type galaxies depends on the redshift, lower galaxies showing a steeper average profile. As deflectors with a steeper profile tend to have a higher lensing cross section, this would favour the deflector redshift distribution to be shifted towards lower redshift.

These effects, along with the redshift dependence of the amplification bias will have to be considered in the analysis of the final XXL sample.

\section*{Acknowledgments}
The authors would like to acknowledge the \textit{Communaut\'e Fran\c caise de Belgique} and the \textit{Actions de Recherche Concert\'ees de l'Acad\'emie universitaire Wallonie-Europe} for their funding during the present research.

\bibliographystyle{mn2e} 


\bsp

\label{lastpage}

\end{document}